\documentclass[epj]{svjour}
\usepackage{latexsym}
\usepackage{graphics}
\usepackage{amssymb}
\usepackage{amsmath}
\usepackage{widetext}
\usepackage{tablefootnote}
\usepackage{xcolor}

\begin{document}
\title{Production of ${^{180\rm{m}}}$Hf in photoproton reaction \\ ${^{181}}$Ta$(\gamma,p)$  at  energy $E_{\rm{\gamma max}}$ = 35--95~MeV}
\author{I.S. Timchenko\inst{1,}\inst{2,}\thanks{\emph{Corresponding author:}\\
		iryna.timchenko@savba.sk;\\  timchenko@kipt.kharkov.ua}%
\and O.S. Deiev\inst{2}
\and S.N. Olejnik\inst{2}
\and S.M. Potin\inst{2}
\and L.P. Korda\inst{2}
\and V.A. Kushnir\inst{2}
\and V.V. Mytrochenko\inst{2,}\inst{3} 
\and S.A.~Perezhogin\inst{2} 
\and A. Herz\'{a}\v{n}\inst{1}  
}                     
%
%
\institute{Institute of Physics, Slovak Academy of Sciences, SK-84511 Bratislava, Slovakia
\and NSC "Kharkov Institute of Physics and Technology", National Academy of Sciences of Ukraine, 1 Academichna Str., Kharkiv, 61108, Ukraine
\and CNRS/IJCLAB, 15 Rue Georges Clemenceau Str., Orsay, 91400, France
}%
\date{Received: date / Revised version: date}
%
\abstract{
The production of the $^{180\rm{m}}\rm{Hf}$ nuclei in the photoproton reaction ${^{181}\rm{Ta}}(\gamma,p)$ was studied at end-point bremsstrahlung energies $E_{\rm{\gamma max}}$ = 35--95~MeV. The experiment was performed at the electron linear accelerator LUE-40 NSC KIPT with the use of the $\gamma$ activation and off-line $\gamma$-ray spectroscopy. 
	The experimental values of the bremsstrahlung flux-averaged cross-sections $\langle{\sigma(E_{\rm{\gamma max}})}\rangle_{\rm{m}}$ for the  ${^{181}\rm{Ta}}(\gamma,p)^{180\rm{m}}\rm{Hf}$ reaction were determined, and at $E_{\rm{\gamma max}} > 55$~MeV obtained for the first time. 
	\\The measured values, also as the literature data, are significantly exceed the theoretical flux-averaged cross-sections $\langle{\sigma(E_{\rm{\gamma max}})}\rangle_{\rm{th}}$. The  $\langle{\sigma(E_{\rm{\gamma max}})}\rangle_{\rm{th}}$ values were calculated using the cross-section $\sigma(E)$ computed with the TALYS1.95 code for six different level density models.  \\
	A comparative analysis of the calculated total cross-sections for the reactions ${^{181}\rm{Ta}}(\gamma,p)^{180}\rm{Hf}$ and ${^{181}\rm{Ta}}(\gamma,n)^{180}\rm{Ta}$ was performed. It was shown that the photoproton $(\gamma,p)$ to photoneutron $(\gamma,n)$ strength ratio is consistent with the estimates based on the isospin selection rules and the value from the $(e,e'p)$ experiment.
} 
\authorrunning{I.S. Timchenko, O.S. Deiev, S.N. Olejnik, S.M. Potin, ...}
\titlerunning{Production of hafnium ${^{180\rm{m}}}$Hf in photoproton reaction ...}
\maketitle

\section{Introduction}\label{intro}

Experimental data on cross-sections for photonuclear reactions are important for many fields of science and technology. These data are necessary for traditional studies of the Giant Dipole Resonance (GDR), and mechanisms of its excitation and decay including competition between statistical and direct processes in decay channels, GDR configurational and isospin splitting, sum rule exhaustion, etc. The cross-sections for photonuclear reactions are also widely used in various applications, primarily in astrophysics \cite{astro}, medicine \cite{med}, design of fast reactors \cite{FR} and accelerator driven sub-critical systems \cite{ADS1,ADS2}. Data on the cross-sections can be found in the comprehensive Atlases \cite{1,2}, and these experimental results are included in the international digital databases EXFOR \cite{3}, ENDF \cite{4}, RIPL \cite{5}, and others. 

It was shown \cite{6,7,8,9,10} that dicrepancies exist in the data on photoneutron cross-sections obtained in different laboratories. This led to work on the analysis of the reliability of previously measured experimental cross-sections \cite{11}, and initiated  new measurements, e.g.  \cite{12,13,14,15}.

In the case of photoproton reactions, an analysis is also required to establish patterns and criteria for data reliability, as, for example, shown in \cite{16}. 
However, there is a lack of experimental data, especially in the region of nuclei with a mass number $A > 100$ for which the $(\gamma,p)$ reaction yields are strongly suppressed.

Previously, photonuclear reactions ${^{181}\rm{Ta}}(\gamma,{\rm{x}}n)^{180-\rm{x}}\rm{Ta}$ were studied for reactions with $\rm{x} \leq 8$ at photon energies up to 130~MeV \cite{11,15,17,18,19,20,21,22,23,24,25,26,27}. Absolute photoneutron cross-sections $\sigma(E)$ were obtained for reactions with $\rm{x} \leq 4$ on quasi-monoenergetic photon beams \cite{11,19,20,21,22,23,24}. Flux-averaged cross-section $\langle{\sigma(E_{\rm{\gamma max}})}\rangle$ for reactions with a large number of particles in the outlet channel were determined using intense beams of bremsstrahlung $\gamma$ rays \cite{15,25,26,27}.

At the same time, according to the EXFOR nuclear database \cite{3}, photonuclear reactions on $^{181}\rm{Ta}$ with the yields of charged particles were studied only in a few works \cite{27,28,29}  due to low cross-section of such reactions. Estimates made in the TALYS1.95 code \cite{30} for the reaction ${^{181}\rm{Ta}}(\gamma,p)^{180\rm{m}}\rm{Hf}$ using the Generalized superfluid level density model ($LD$3)  give a value of cross-section $\sigma(E)$ $\approx$ 0.0136 mb at the maximum ($E$ $\approx$ 27~MeV). For comparison, in the case of the ${^{181}\rm{Ta}}(\gamma,n)$ reaction the experimental values of $\sigma(E)$ are about 400 mb at the energy of GDR maximum \cite{11}.  Note that values of $\sigma(E)$ of the  ${^{181}\rm{Ta}}(\gamma,p)^{180\rm{m}}\rm{Hf}$ reaction recalculated to flux-averaged cross-section $\langle{\sigma(E_{\rm{\gamma max}})}\rangle$ using GEANT4.9.2 \cite{31} decreased to 3.0--4.2~$\mu$b in the energy range 35--95~MeV.

Previously, the experimental yield for reaction \\ ${^{181}\rm{Ta}}(\gamma,p)^{180\rm{m}}\rm{Hf}$ relative to ${^{181}\rm{Ta}}(\gamma,n)^{180}\rm{Ta}$ 
 were obtained at end-point bremsstrahlung energy $E_{\rm{\gamma max}}$ = 67.7 MeV and was found value $(5 \pm 1) \times 10^{-4}$ \cite{27}. This relative yield 
 was compared with the calculation in TALYS, which gives a value of $3 \times 10^{-5}$. Since the calculation of the flux-averaged cross-section for the  ${^{181}\rm{Ta}}(\gamma,n)^{180}\rm{Ta}$ reaction in the TALYS1.95 code agrees well 
with the experimental data \cite{15,25} in a wide energy range, the observed discrepancy in the relative yield must be due 
 to the differences between the experimental and calculated cross-sections for the production of the ${^{180\rm{m}}\rm{Hf}}$ nuclei. 
 
 The flux-averaged yields for the ${^{181}\rm{Ta}}(\gamma,p)^{180\rm{m}}\rm{Hf}$ reaction were studied at $E_{\rm{\gamma max}}$ = 20, 40, and 55~MeV in \cite{28}. The comparison of experimental data with the TALYS1.9 code showed a significant discrepancy with the calculation.



In this work, we studied the production of $^{180\rm{m}}\rm{Hf}$ in a photoproton reaction 
 by means of $\gamma$-ray spectroscopy. The bremsstrahlung flux-averaged cross-section $\langle{\sigma(E_{\rm{\gamma max}})}\rangle_{\rm{m}}$ was determined in the range of end-point bremsstrahlung energy $E_{\rm{\gamma max}}$ = 35--95~MeV. Also, theoretical calculations using the TALYS1.95 code  were performed. 

A comparative analysis of the calculated total cross-sections for the reactions ${^{181}\rm{Ta}}(\gamma,p)^{180}\rm{Hf}$ and \\ ${^{181}\rm{Ta}}(\gamma,n)^{180}\rm{Ta}$  was done. The photoproton $(\gamma,p)$ to photoneutron $(\gamma,n)$ strength ratio was calculated according to the isospin selection rules. This data was compared with calculations in TALYS1.95, and with the experimental value obtained using the cross-sections from the $(e,e'p)$ experiment \cite{Shoda46}.

\section{Experimental procedure and flux-averaged cross-sections determination}\label{sec2}
\subsection{Experimental setup and method}
\label{subsec21}

The experiment to study the production of the $^{180\rm{m}}\rm{Hf}$ nuclei in photoproton reaction ${^{181}\rm{Ta}}(\gamma,p)    $ was carried out using the method of measuring the residual $\gamma$-activity of an irradiated sample. This technique enables us to obtain simultaneously the data from different channels of photonuclear reactions, e.g. $(\gamma,p)$, $(\gamma,n)$, $(\gamma,2n)$ etc., for example, \cite{32,33,34,35}.  

The experiment was performed at the National Science Center "Kharkov Institute of Physics and Technology" (NSC KIPT), Ukraine, employing the electron linear accelerator LUE-40  \cite{36,37}. To generate bremsstrahlung $\gamma$ quanta, electrons with initial energy  $E_e$ impinged on a converter made of a natural tantalum plate with transverse dimensions of $20 \times 20$~mm and a thickness of 1.05~mm. The flux of bremsstrahlung  $\gamma$ quanta was cleaned from electrons using an Al absorber of cylindrical shape with a diameter of 100~mm and a length of 150~mm. 

\begin{figure}[h]%
	\centering
	\resizebox{0.5\textwidth}{!}{%
		\includegraphics{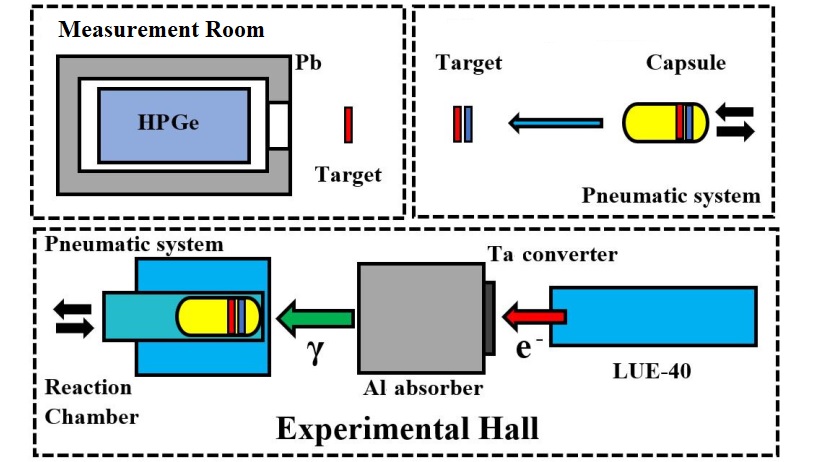}}
	\caption{The schematic block diagram of the experimental setup. The upper part shows the measurement room, where the exposed target (red colour) and the Mo target-monitor (blue colour) are extracted from the capsule and placed one by one to the HPGe detector for induced $\gamma$-activity measurements. The lower part shows the accelerator LUE-40, Ta converter, Al absorber, and the exposure reaction chamber.}\label{fig1}
\end{figure}

 There were two types of targets in the experiment. The ${^{\rm nat}}$Ta target was used to investigate the production of the  $^{180\rm{m}}\rm{Hf}$ nuclei, while the purpose of the ${^{\rm nat}}$Mo target-monitor was to control a $\gamma$-ray flux. The ${^{\rm nat}}$Ta and ${^{\rm nat}}$Mo targets had a shape of a disk with a diameter of 8~mm. Their thicknesses and masses were 50 $\mu$m, $\sim$43~mg and 100~$\mu$m, $\sim$60~mg, respectively. Both targets were simultaneously placed in a thin aluminum capsule and transported to/out of the irradiation site using a pneumatic transport system. After that, irradiated samples were taken to the measurement room where, after their  removal from the Al capsule, the residual activities were measured. The cooling time for the Ta target, taking into account the transfer and removal of the target from the capsule, was no more than 2--3~min. The irradiation time and the duration of measuring the residual $\gamma$ activity spectrum  were both 30~min long. The scheme of the experiment is shown in Fig.~\ref{fig1}.

The induced $\gamma$-activity of the irradiated targets was measured by the semiconductor high-purity germanium (HPGe) detector, model Canberra GC-2018 with the energy resolution (FWHM) of 0.8 and 1.8~keV at 122 and 1332.5~keV, respectively. Its detection efficiency, $\varepsilon$, at \\ 1332.5~keV was 20\% relative to the NaI(Tl) scintillator, 3~inches in diameter and 3~inches in thickness. Calibration of the detection efficiency was done by using a set of $\gamma$-ray radiation sources: $^{22}$Na, $^{60}$Co, $^{133}$Ba, $^{137}$Cs, $^{152}$Eu, $^{241}$Am. The numerical value of $\varepsilon$ was determined for various $\gamma$-ray energies using the analytical curve in the form ${\rm{ln}}\varepsilon = \sum\limits_{i=1}^n a_i ({\rm{ln}} E_{\gamma})^i$ proposed in \cite{38}. 

At the end-point bremsstrahlung energies $E_{\rm{\gamma max}}$ = 60.4 and 80.5~MeV, additional measurements of the \\${^{181}\rm{Ta}}(\gamma,p)^{180\rm{m}}\rm{Hf}$ reaction cross-sections  were performed at the different experimental setup \cite{39,39a}. In this case, a $^{\rm{nat}}$Ta converter foil with a thickness of 100~$\mu$m, and a bending magnet to clean the bremsstrahlung $\gamma$-flux from electrons were used.

The bremsstrahlung spectra of electrons were simulated using the GEANT4.9.2 toolkit \cite{31}. In the simulation, the real relative position of the target, Ta converter, Al absorber, and elements of the experimental equipment, as well as the spatial and energy distribution of the electron beam were used as the input parameters.

The monitoring of the calculated bremsstrahlung $\gamma$ flux was performed with the use of the $^{100}{\rm{Mo}}(\gamma,n)^{99}\rm{Mo}$ reaction yield. For this purpose, the experimentally obtained flux-averaged cross-sections were compared with the theoretical values. To determine the experimental\\ $\langle{\sigma(E_{\rm{\gamma max}})}\rangle$  values, we have used the number of counts under the $\gamma$-ray peak at $E_{\gamma}$ = 739.50~keV  with the intensity $I_{\rm \gamma}$ = 12.13\%  \cite{40}. 
The theoretical values of the flux-averaged cross-section $\langle{\sigma(E_{\rm{\gamma max}})}\rangle_{\rm{th}}$  were calculated using the cross-sections $\sigma(E)$ from the TALYS1.95 code \cite{30}. The resulting normalization coefficients \\$k_{\rm{mo}}$ = $\langle{\sigma(E_{\rm{\gamma max}})}\rangle_{\rm{th}}$/$\langle{\sigma(E_{\rm{\gamma max}})}\rangle$ were used to normalize the cross-sections of the reaction under study. More details about the monitoring procedure can be found in \cite{35,41,42}.

The irradiated Ta converter and Al absorber generate  neutrons that can trigger the reaction $^{100}{\rm{Mo}}(n,2n)^{99}\rm{Mo}$. To evaluate also this option, energy spectra of neutrons above the threshold energies  were calculated using the GEANT4.9.2, similarly to \cite{43}.
The contribution of the $^{100}{\rm{Mo}}(n,2n)^{99}\rm{Mo}$ reaction to the value of the induced activity of the $^{99}$Mo nucleus has been estimated and it has been shown that this contribution is negligible compared to the contribution of  $^{100}{\rm{Mo}}(\gamma,n)^{99}\rm{Mo}$. The contribution of the reaction $^{100}{\rm{Mo}}(\gamma,p)^{99}\rm{Nb}$, $^{99}\rm{Nb}\xrightarrow{\beta^-}$$^{99}\rm{Mo}$ is also negligible.

\subsection{Calculation of the flux-averaged cross-sections}
\label{subsec22}

The values of the theoretical cross-section $\sigma(E)$ computed with the TALYS1.95 code \cite{30} were averaged over the bremsstrahlung $\gamma$-flux $W(E,E_{\rm{\gamma max}})$ from the threshold energy $E_{\rm{thr}}$ of the reaction under study to the end-point bremsstrahlung energy $E_{\rm{\gamma max}}$. As a result of this procedure, the flux-averaged cross-sections $\langle{\sigma(E_{\rm{\gamma max}})}\rangle_{\rm th}$ were calculated using the equation: 

\begin{equation}\label{form1}
	\langle{\sigma(E_{\rm{\gamma max}})}\rangle_{\rm th} = 
	{{\rm{\Phi}}^{-1}(E_{\rm{\gamma max}}) \int\limits_{E_{\rm{thr}}}^{E_{\rm{\gamma max}}}\sigma(E)  W(E,E_{\rm{\gamma max}})dE},
\end{equation}
where ${\rm{\Phi}}(E_{\rm{\gamma max}}) = {\int\limits_{E_{\rm{thr}}}^{E_{\rm{\gamma max}}}W(E,E_{\rm{\gamma max}})dE}$  is the integrated bremsstrahlung $\gamma$-flux.

Theoretical flux-averaged cross-sections were compared with those measured in the experiment and calculated as follows: 
\begin{equation}
		\langle{\sigma(E_{\rm{\gamma max}})}\rangle = 
		\frac{\lambda \triangle A  {\rm{\Phi}}^{-1}(E_{\rm{\gamma max}})}{N_x I_{\gamma} \ \varepsilon (1-e^{-\lambda t_{\rm{irr}}})e^{-\lambda t_{\rm{cool}}}(1-e^{-\lambda t_{\rm{meas}}})},
		\label{form2}
\end{equation}
where $\triangle A$ is the  number of counts in the full absorption $\gamma$-ray peak; $\lambda$ denotes the decay constant \mbox{($\rm{ln}2/\textit{T}_{1/2}$)}; $T_{1/2}$ is the half-life of the nucleus; $N_x$ is the number of target atoms; $I_{\gamma}$ is the intensity of the analyzed $\gamma$ ray; 
$\varepsilon$ is the detection efficiency at the energy of analyzed $\gamma$ ray;
 $t_{\rm{irr}}$, $t_{\rm{cool}}$ and $t_{\rm{meas}}$ are the irradiation time, cooling time and measurement time, respectively.
A more detailed description of all the calculation procedures necessary for the determination of $\langle{\sigma(E_{\rm{\gamma max}})}\rangle$ can be found in \cite{15,35}.

To determine the experimental $\langle{\sigma(E_{\rm{\gamma max}})}\rangle$, we used the parameter values of the reaction ${^{181}\rm{Ta}}(\gamma,p)^{180\rm{m}}\rm{Hf}$ listed in Table~\ref{tab1}. 
In addition,  parameter values  for \\  ${^{181}\rm{Ta}}(\gamma,6n)^{175}\rm{Ta}$,  ${^{181}\rm{Ta}}(\gamma,p)^{180\rm{m}}\rm{Hf}$,  ${^{181}\rm{Ta}}(\gamma,n)^{180\rm{g}}\rm{Ta}$,\\ ${^{100}\rm{Mo}}(\gamma,n)^{99}\rm{Mo}$ reactions are listed as well.

\begin{table*}[ht]
	\begin{center}
		\caption{ Spectroscopic data of the products of different reactions adopted from \cite{40}: spin $J$, parity $\pi$, half-life $T_{1/2}$ of the reaction products; $E_{\gamma}$ and $I_{\gamma}$ are the energies of the $\gamma$-ray transitions and their intensities, respectively. $E_{\rm{thr}}$ denotes threshold energy of the reactions.}\label{tab1}  
		\begin{tabular}{cccccc}
			\hline \hline\noalign{\smallskip}
			Nuclear reaction & $E_{\rm{thr}}$,~MeV & $J^{\pi} $& $T_{1/2}$, h & $E_{\gamma}$,~keV & $I_{\gamma}$, \% \\ \noalign{\smallskip}\hline\noalign{\smallskip}
			${^{181}\rm{Ta}}(\gamma,p)^{180\rm{m}}\rm{Hf}$  & 7.09 & $8^-$ & 5.5 $\pm$ 0.1 & \begin{tabular}{c}443.09  $\pm$ 0.04 \\500.64  $\pm$ 0.18
			\end{tabular}
			& \begin{tabular}{c} 81.9 $\pm$ 0.9 \\ 14.3 $\pm$ 0.3
			\end{tabular} \\ \noalign{\smallskip}\hline\noalign{\smallskip}		
			${^{181}\rm{Ta}}(\gamma,n)^{180\rm{g}}\rm{Ta}$  & 7.58 & $1^+$ & 8.152 $\pm$ 0.006 & 103.557  $\pm$ 0.007 & 0.81 $\pm$ 0.16 \\ \noalign{\smallskip}\hline\noalign{\smallskip}
			${^{181}\rm{Ta}}(\gamma,6n)^{175}\rm{Ta}$  & 44.46 & 7/2$^+$ & 10.5 $\pm$ 0.2 & \begin{tabular}{c} 348.5  $\pm$ 0.5 \\ 436.4  $\pm$ 0.7 \\443.3  $\pm$ 0.7 
			\end{tabular} & \begin{tabular}{c} 12.0  $\pm$ 0.6 \\ 3.8  $\pm$ 0.2 \\0.14  $\pm$ 0.04
		\end{tabular} \\ \noalign{\smallskip}\hline\noalign{\smallskip}	
			$^{100}{\rm{Mo}}(\gamma,n)^{99}\rm{Mo}$ & 8.29 & 1/2$^+$ & $65.94 \pm 0.01$ &
			739.50  $\pm$ 0.02 &       	$12.13 \pm 0.12$ \\	    \noalign{\smallskip}\hline\hline
		\end{tabular}	
	\end{center}
\end{table*}

Note that if the reaction product has  known isomeric state, the total flux-averaged cross-section $\langle{\sigma(E_{\rm{\gamma max}})}\rangle_{\rm{tot}}$ is calculated as the sum of cross-sections for the ground state $\langle{\sigma(E_{\rm{\gamma max}})}\rangle_{\rm{g}}$ and isomeric state $\langle{\sigma(E_{\rm{\gamma max}})}\rangle_{\rm{m}}$, respectively.

\subsection{Experimental accuracy of flux-averaged cross-sections}
\label{subsec23}

     The uncertainty of measured flux-averaged cross-sections was determined as a square root of the quadratic sum of statistical and systematic errors. The statistical error in the observed $\gamma$-activity is mainly due to statistics in the full absorption peak of the corresponding $\gamma$-ray, which varies within 2 to 9\%. 
     The measured $\triangle A$ value of the investigated $\gamma$ ray depends on the detection efficiency,  half-life, and the intensity $I_{\rm \gamma}$. The background is generally governed by the contribution from the Compton scattering of the emitted  $\gamma$ rays.

The systematical errors are due to the following uncertainties of the: 

1.	exposure time and the electron current  $\sim$0.5\%; 

2. $\gamma$-ray detection efficiency of the detector -- 2--3\%.
 The error is larger at $E_{\rm \gamma}$ = 50--200~keV, this being due to a small number of calibration data points in this energy range and the convoluted shape of the efficiency curve;  

3. the half-life  $T_{1/2}$ of the reaction products and the intensity $I_{\rm \gamma}$ of the analyzed $\gamma$ rays;

 4. normalization of the experimental data to the yield of the monitoring reaction $^{100}{\rm{Mo}}(\gamma,n)^{99}\rm{Mo}$ made up 6~\%. It should be noted that the systematic error in yield monitoring of the $^{100}{\rm{Mo}}(\gamma,n)^{99}\rm{Mo}$ reaction stems from three errors, each reaching up to 1\%. These are the statistical error in the determination of the number of counts under the $\gamma$-ray peak used for normalization, the uncertainty in the isotopic composition of natural molybdenum and in the intensity $I_{\rm \gamma}$ used. In our calculations, we have used the percentage value of $^{100}\rm{Mo}$ isotope abundance equal to 9.63\% \cite{31}.
 
The total uncertainties of the measured flux-averaged cross-sections are given in Fig.~\ref{fig4} and Table~\ref{tab2}.

\section{Results and discussion}\label{sec3}

\subsection{Experimental values of bremsstrahlung flux-averaged cross-section $\langle{\sigma(E_{\rm{\gamma max}})}\rangle_{\rm{m}}$}
\label{subsec31}


\begin{figure*}[]
	\resizebox{1.0\textwidth}{!}{%
		\includegraphics{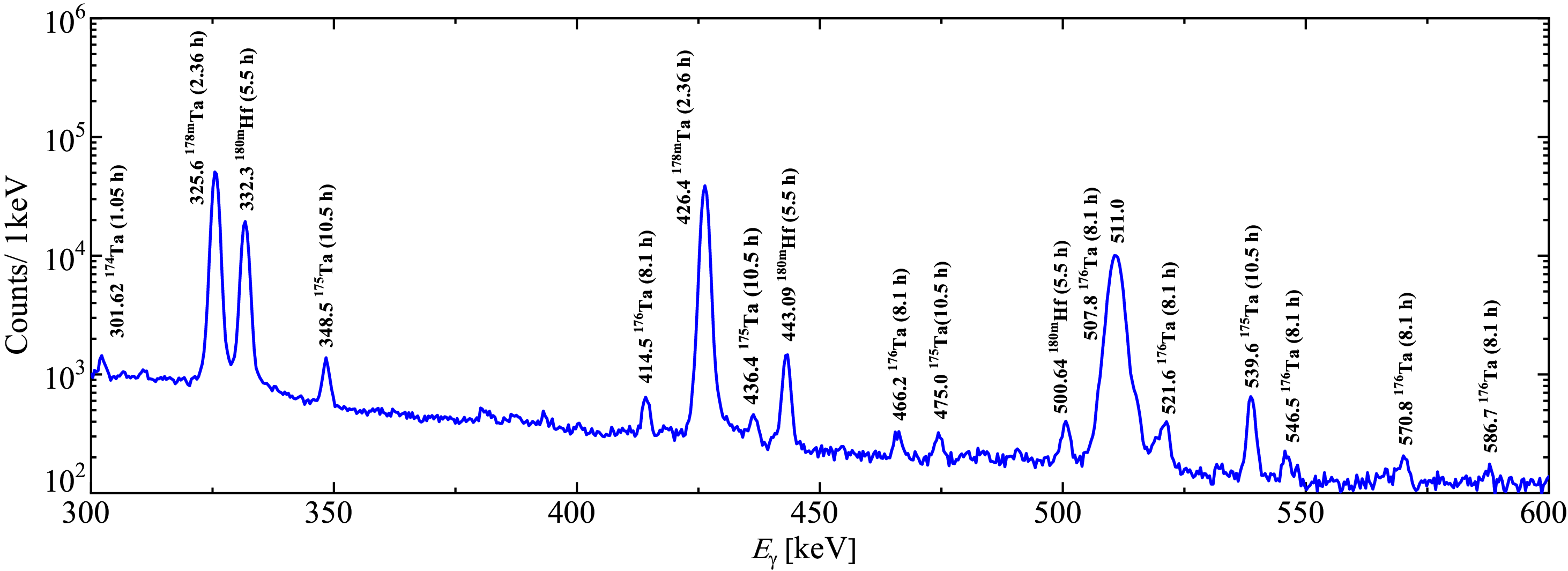}}
	\caption{Energy spectrum of $\gamma$ rays  measured by the HPGe detector from the 42.598~mg $^{181}$Ta target after bremsstrahlung flux exposure time of 30~min with $E_{\rm{\gamma max}}$ = 80.2~MeV. Spectrum fragment ranging from 300 to 600~keV is shown.}
	\label{fig2}
\end{figure*}

To calculate the experimental flux-averaged cross-sections for the ${^{181}\rm{Ta}}(\gamma,p)^{180\rm{m}}\rm{Hf}$ reaction, two $\gamma$-ray transitions with energies of 443.09 and 500.64~keV were used, see Fig. 2. The 443.09~keV $\gamma$-ray peak is preferable for use because of the much higher intensity. However, there is a $\gamma$-ray with a similar energy of 443.3~keV corresponding to the ${^{175}\rm{Ta}}$ nucleus, which is product of the ${^{181}\rm{Ta}}(\gamma,6n)$ reaction. Because of the almost identical transition energies, the two peaks overlap, thus artificially increasing the intensity of the 443.09 keV peak. This fact had to be taken into account in the analysis.
To estimate the magnitude of this contribution, we used the 348.5 and 436.4~keV  $\gamma$-rays with the intensity $I_{\gamma} =$ 12.0\% and 3.8\%, respectively, which correspond to the $^{175}\rm{Ta}$ nucleus. The obtained $\triangle A$  in these $\gamma$-ray peaks, taking into account the detection efficiency $\varepsilon$, were recalculated to values of the activity of the $^{175}\rm{Ta}$ nucleus by the 443.3~keV $\gamma$-ray. 
 The resulting contribution did not exceed 1\%. 



It should be noted that the calculation of the contribution of the competing reaction cannot be performed with an accuracy better than 29\%, which is associated with a large error in the intensity $I_{\rm{\gamma}}$ for the 443.3~keV transition, see Table~\ref{tab1}. Since the yield of the $^{181}{\rm{Ta}}(\gamma,6n)^{175}\rm{Ta}$ reaction is $< 1$ \% of the yield of the $^{181}{\rm{Ta}}(\gamma,p)^{180\rm{m}}\rm{Hf}$ reaction, even so large error has a negligible effect on the final result.

\begin{figure}[t]
	\resizebox{0.49\textwidth}{!}{%
		\includegraphics{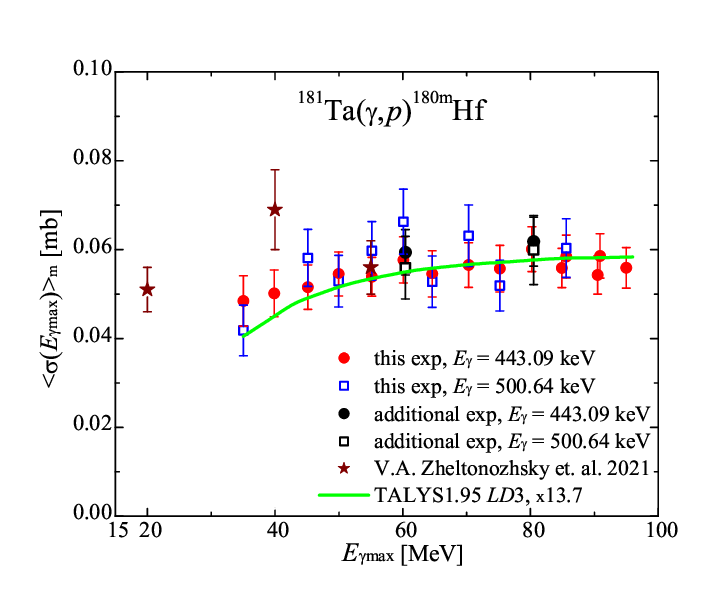}}
	\caption{Bremsstrahlung flux-averaged cross-sections $\langle{\sigma(E_{\rm{\gamma max}})}\rangle_{\rm{m}}$ of the $^{181}{\rm{Ta}}(\gamma,p)^{180\rm{m}}\rm{Hf}$ reaction. Red and black circles -- our experimental data for $\gamma$-ray with $E_{\rm{\gamma}}$ = 443.09~keV, blue and black empty squares -- for $E_{\rm{\gamma}}$ = 500.64~keV, brown stars -- data taken from \cite{28}. The additional measurements at $E_{\rm{\gamma max}}$ = 60.4 and 80.5~MeV are denoted as black full circles and black empty squares. Curve -- calculation using TALYS1.95 code for model $LD$3, upscaled by a factor of 13.7.}
	\label{fig4}
	\vspace{1ex}
\end{figure}

The experimental values of the bremsstrahlung flux-averaged cross-sections $\langle{\sigma(E_{\rm{\gamma max}})}\rangle_{\rm{m}}$ for the reaction \\ $^{181}{\rm{Ta}}(\gamma,p)^{180\rm{m}}\rm{Hf}$ obtained in this work at the end-point bremsstrahlung energies $E_{\rm{\gamma max}}$ = 35--95~MeV are shown in Fig.~\ref{fig4} and Table~\ref{tab2}. As can be seen, within experimental uncertainties, the obtained cross-sections for the 443.09 and 500.64 keV $\gamma$-ray transitions are in agreement.

The values of $\langle{\sigma(E_{\rm{\gamma max}})}\rangle_{\rm{m}}$ obtained in additional measurements at $E_{\rm{\gamma max}}$ = 60.4 and 80.5~MeV are in good agreement with all massive data (see Fig.~\ref{fig4}).
 
The obtained experimental results were compared with the data from literature \cite{28}, which received at $E_{\rm{\gamma max}}$ = 20, 40 and 55 MeV. From Fig.~\ref{fig4} it can be seen, that the experimental cross-sections are in good agreement at $E_{\rm{\gamma max}}$ = 55 MeV, and don't agree at lower energies. 
 
One more comparison was made with the data published in \cite{27}. There, the experimental $^{181}{\rm{Ta}}(\gamma,p)^{180\rm{m}}\rm{Hf}$ reaction yield relative to the $^{181}{\rm{Ta}}(\gamma,n)^{180\rm{g}}\rm{Ta}$ reaction yield was obtained at $E_{\rm{\gamma max}}$ = 67.7~MeV, and equal to $(5 \pm 1)	\times10^{-4}$.
For the comparison, the flux-averaged cross-sections determined in this work in the energy range $E_{\rm{\gamma max}}$ = 60--80~MeV were approximated and the value $\langle{\sigma(E_{\rm{\gamma max}})}\rangle_{\rm{m}}$ = $0.057 \pm 0.005$ mb at 67.7~MeV was calculated. To obtain the relative yield, the experimental values of the flux-averaged cross-section of the $^{181}{\rm{Ta}}(\gamma,n)^{180\rm{g}}\rm{Ta}$ reaction from \cite{15,25} were used, taking into account the difference in bremsstrahlung $\gamma$-flux due to the difference in the reaction thresholds $E_{\rm{thr}}(\gamma, n)$ = 7.58~MeV and $E_{\rm{thr}}(\gamma, p)$ = 7.09~MeV. We get $(7.2 \pm 1.4) \times 10^{-4}$ for the relative yield. Within  experimental uncertainties, it agrees with the value from \cite{27}. 

 \begin{table}[h]
	\caption{\label{tab2}  Experimental flux-averaged cross-sections $\langle{\sigma(E_{\rm{\gamma max}})}\rangle_{\rm{m}}$ of the $^{181}{\rm{Ta}}(\gamma,p)^{180\rm{m}}\rm{Hf}$ reaction (data for $E_{\rm{\gamma}}$ = 443.09~keV).}
	\centering
		\begin{tabular}{cc}
 				\hline\hline\noalign{\smallskip}
		 $E_{\rm{\gamma max}}$,~MeV & $\;\;\;\;\;\;\langle{\sigma(E_{\rm{\gamma max}})}\rangle_{\rm{m}}$, $\mu$b   \\ \noalign{\smallskip}\hline\noalign{\smallskip}	
		 35.1 & 48.4 $\pm$ 5.7  \\ 
		 39.9 & 50.2 $\pm$ 5.3  \\ 
		 45.1 & 51.6 $\pm$ 5.0  \\ 
		 50.0 & 54.5 $\pm$ 5.0  \\ 
		 55.2 & 53.9 $\pm$ 4.4  \\ 
		 60.1 & 57.7 $\pm$ 5.2  \\ 
		 		 60.4* & 59.4 $\pm$ 5.1  \\ 
		 64.6 & 54.6 $\pm$ 5.2  \\ 
		 70.3 & 56.5 $\pm$ 5.0  \\ 
		 75.2 & 55.7 $\pm$ 5.2  \\ 
		  80.2 & 60.1 $\pm$ 5.0  \\ 
		 		 80.5* & 61.8 $\pm$ 5.5  \\ 
		 84.9 & 55.9 $\pm$ 4.4  \\ 
		 85.6 & 58.5 $\pm$ 4.8  \\ 
		 90.5 & 54.3 $\pm$ 4.3  \\ 
		 90.9 & 58.6 $\pm$ 5.0  \\ 
		 95.0 & 55.9 $\pm$ 4.6  \\ \noalign{\smallskip}\hline\hline	  	
		\end{tabular} \\ 
\footnotesize{* The data were obtained in an additional experiment carried out on the setup described in \cite{39}.}
	\end{table}

 \subsection{ Calculated $\sigma(E)$ and $\langle{\sigma(E_{\rm{\gamma max}})}\rangle$ cross-sections}\label{subsec32}

The theoretical values of total and partial (metastable, ground) cross-sections $\sigma(E)$ for the ${^{181}\rm{Ta}}(\gamma,p)^{180}\rm{Hf}$  reaction with monochromatic photons were calculated using the \\ TALYS1.95 code \cite{30}. The calculations were performed for six different level density models denoted as $LD$ 1-6. There are three phenomenological level density models and three options for microscopic level densities:

$LD 1$: Constant temperature + Fermi gas model, introduced by Gilbert and Cameron \cite{Gilbert}.
 In this model, the excitation energy range is divided into a low energy part from $E_0$ up to a matching energy $E_{\rm{M}}$, where the so-called constant temperature law applies and a high energy part above, where the Fermi gas model applies. 

$LD 2$: Back-shifted Fermi gas model \cite{Back-shifted}, 
where the pairing energy is treated as an adjustable parameter and the Fermi gas expression is used down to $E_0$.

$LD 3$: Generalized superfluid model (GSM) \cite{Ignatyuk1,Ignatyuk2}. 
The model takes superconductive pairing correlations into account according to the Bardeen-Cooper-Schrieffer theory. 

$LD 4$: Microscopic level densities (Skyrme force) from Goriely’s tables \cite{Goriely}. 
Using this model allows reading tables of microscopic level densities from RIPL database \cite{5}. These tables were computed by S. Gorielyon based on Hartree-Fock calculations for excitation energies up to 150~MeV and for spin values up to $I$ = 30. 

$LD 5$: Microscopic level densities (Skyrme force) from Hilaire’s combinatorial tables \cite{Hilaire}. 
The combinatorial model includes a detailed microscopic calculation of the intrinsic state density and collective enhancement. The only phenomenological aspect of the model is a simple damping function for the transition from spherical to deformed.  

$LD 6$: Microscopic level densities based on temperature-dependent Hartree-
Fock-Bogoliubov calculations using the Gogny force \cite{Gorny} from Hilaire’s combinatorial tables.

\begin{figure*}[]
	\resizebox{1.\textwidth}{!}{%
		\includegraphics{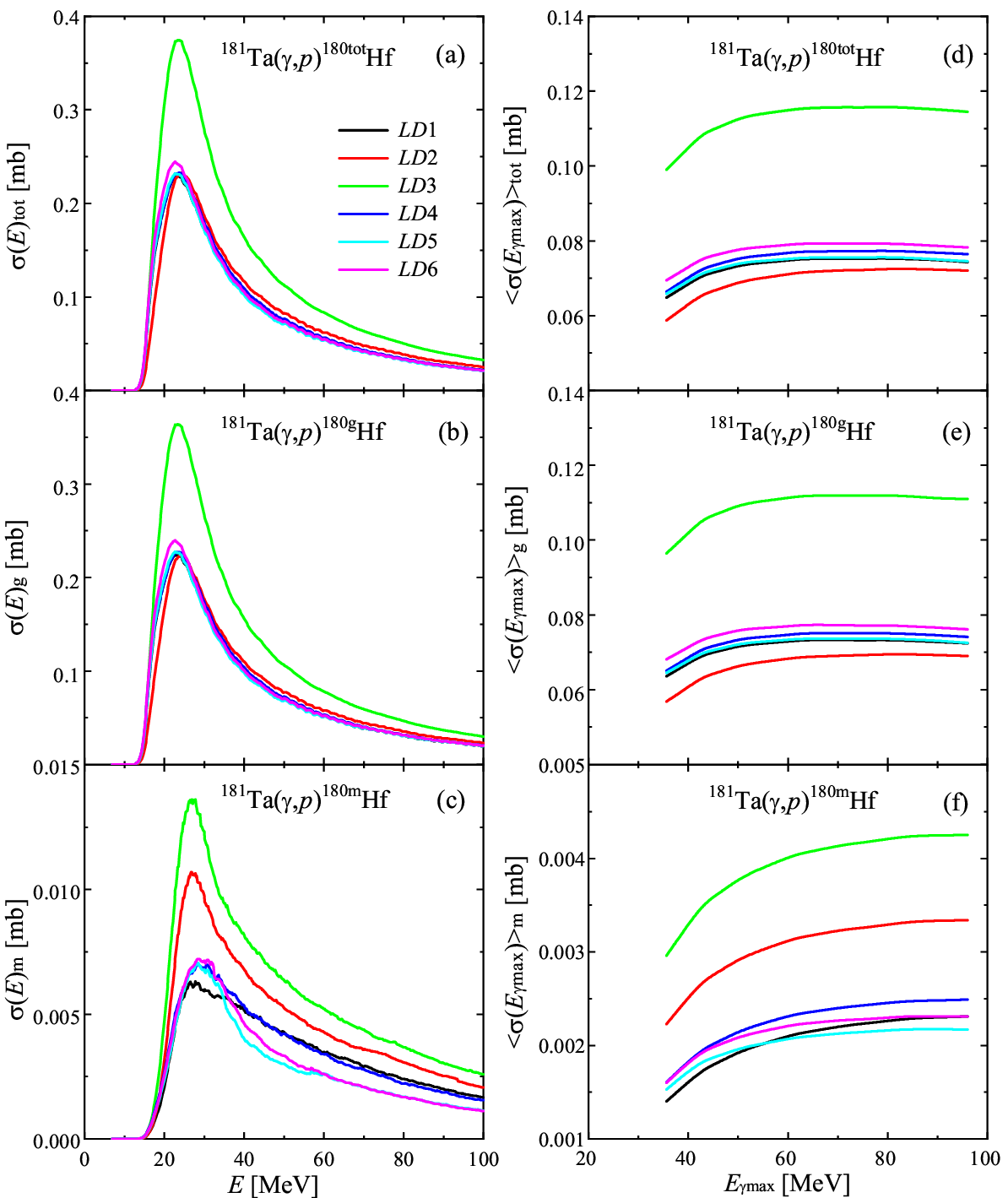}}
	\caption{Theoretical values of the $^{181}{\rm{Ta}}(\gamma,p)^{180}\rm{Hf}$ reaction cross-sections calculated using the TALYS1.95 code for the level density models \textit{LD}~1--6. The absolute cross-sections $\sigma(E)$ for total, ground and metastable states for monochromatic photons are shown in panels (a)-(c), respectively. The bremsstrahlung flux-averaged cross-sections $\langle{\sigma(E_{\rm{\gamma max}})}\rangle$ for total, ground and metastable states are shown in panels (d)-(f), respectively.}
	\label{fig3}
\end{figure*}

Results of the calculations of the total, metastable and ground state cross-sections $\sigma(E)$ for the ${^{181}\rm{Ta}}(\gamma,p)^{180}\rm{Hf}$ reaction are shown in Figs.~\ref{fig3}(a)-(c). The bremsstrahlung flux-averaged cross-sections $\langle{\sigma(E_{\rm{\gamma max}})}\rangle_{\rm tot}$, $\langle{\sigma(E_{\rm{\gamma max}})}\rangle_{\rm g}$, and  $\langle{\sigma(E_{\rm{\gamma max}})}\rangle_{\rm m}$ are presented in Figs.~\ref{fig3}(d)-(f). 

According to the calculated cross-sections $\sigma(E)$ and $\langle{\sigma(E_{\rm{\gamma max}})}\rangle$, formation of the $^{180}\rm{Hf}$ nucleus in the metastable state is strongly suppressed. The contribution of the metastable state in the total flux-averaged cross-section does not exceed 5\% for all level density models.

The  largest values of the $\langle{\sigma(E_{\rm{\gamma max}})}\rangle_{\rm m}$ and $\langle{\sigma(E_{\rm{\gamma max}})}\rangle_{\rm g}$ cross-sections are obtained with the $LD$3 level density model (see Fig.~\ref{fig3}). 

\subsection{Comparison of experimental data with theoretical calculations}
\label{subsec:33}

Comparison of obtained experimental cross-sections \\$\langle{\sigma(E_{\rm{\gamma max}})}\rangle_{\rm{m}}$ for the $^{181}{\rm{Ta}}(\gamma,p)^{180\rm{m}}\rm{Hf}$ reaction with the theoretical estimates shows that all calculated variants ($LD$ 1-6) are underestimated.
 Using the least squares method it was found that the best agreement is achieved with the $LD$3 model, where the calculated flux-averaged cross-section are smaller by a factor of 13.7. 
Rescaled result of the $LD$3 model calculation is graphically shown and compared with the measured data in Fig.~\ref{fig4}. In further text, only calculations with the $LD$3 model will be discussed.

 Note that a difference of similar magnitude is also valid for the experimental data from \cite{27,28}. 
 
Due to $^{180{\rm{g}}}\rm{Hf}$ being stable, it was not possible to determine the cross-section for its production in our experiment. We are not aware of any published experimental data on this topic. 
Therefore, only indirect estimate of the total cross-section for the $^{181}{\rm{Ta}}(\gamma,p)^{180}\rm{Hf}$ reaction could be made based on the isospin selection rules \cite{44}.

For photodisintegration in heavy nuclei, the isospin-splitting components of GDR can be approximated by the $(\gamma,n)$ and $(\gamma,p)$ cross-sections, respectively. Let us estimate the strength ratio of the $(\gamma,p)$ to the $(\gamma,n)$ components of GDR using the expression from \cite{45}, which can be written as:
\begin{equation}\label{form3}
	{\int\limits_{0}^{\infty}  \frac{\sigma_{\gamma p}}{E} dE} \Bigg/
	{\int\limits_{0}^{\infty}  \frac{\sigma_{\gamma n}}{E}  dE} = 	
	\frac {1}{T_0} \times \frac {1-1.5 T_0 A^{-2/3}}{1 + 1.5 A^{-2/3}},
\end{equation}
where $\sigma_{\gamma p}$ and $\sigma_{\gamma n}$ are total cross-sections corresponding to $(\gamma,p)$ and $(\gamma,n)$ reactions, $T_0 = (N - Z)/2$ is isospin of the ground state of the nucleus with $N$ neutrons and $Z$ protons, $A = N + Z$. 

For the $^{181}{\rm{Ta}}$ nucleus, $T_0$ = 17.5, and the expected value of the strength ratio is $9.81 \times 10^{-3}$ (calculated using the right part of Eq.~\ref{form3}). At the same time, calculation using weighted integrals up to 100~MeV gives a ratio equal to $2.35 \times 10^{-3}$. 
It was shown in \cite{15} that the theoretical cross-section for the $^{181}{\rm{Ta}}(\gamma,n)$ reaction describes well the experimental results. Therefore, the observed discrepancy between the calculated values is related to the total cross-section for the $^{181}{\rm{Ta}}(\gamma,p)$ reaction. 

As shown before (see Fig.~\ref{fig4}), the experimental data for the flux-averaged cross-sections for the population of the isomeric state significantly exceed the theoretical values. Since the  $\sigma(E)_{\rm m}$ cross-section is small relative to the  $\sigma(E)_{\rm g}$, then taking into account the found factor of 13.7, slightly increases the weighted integrals ratio (left part of Eq.~\ref{form3}), up to $3.47 \times 10^{-3}$. 

	
	It was shown in \cite{45}, that isospin selection rule approach gives an average value for a set of nuclei, while in case of a single nucleus, a deviation can be significant, e.g, for the ${^{208}\rm{Pb}}$ nucleus, the expected theoretical strength ratio $ 2.6 \times 10^{-3}$ is 1.9 times lower than the experimentally measured one; for the ${^{139}\rm{La}}$ nucleus, theory gives a 2.4 times larger value than measurement. 
	
	The calculation carried out (\cite{Shoda46} and references therein) for the ${^{181}\rm{Ta}}$ nucleus using the experimental cross-sections from the $(e,e'p)$ experiment gave the value of the strength ratio $\sim$1.7$ \times 10^{-3}$. The error in this estimate may be large, due to the absence of data for the experimental $(e,e'p)$ cross-section in high-energy region for a better fitting.
	
Thus, the  strength ratio for the ${^{181}\rm{Ta}}$ nucleus, determined to be $3.47 \times 10^{-3}$, lies between the expected estimate according to the isospin selection rules, and the strength ratio obtained using the experimental cross-sections from the $(e,e'p)$ experiment \cite{Shoda46}.

\section{Conclusions}
\label{sec4}

The experimental study of the production of metastable $^{180}\rm{Hf}$ nucleus in the photoproton reaction $^{181}{\rm{Ta}}(\gamma,p)^{180\rm{m}}\rm{Hf}$ was performed at end-point  bremsstrahlung energies $E_{\rm{\gamma max}}$ = 35--95~MeV. The experiment was carried out at the NSC KIPT, Ukraine, with bremsstrahlung beams generated by the electron linear accelerator LUE-40 and using the $\gamma$ activation and off-line $\gamma$-ray spectrometric technique. There were two different experimental setups used, the results of which are in good agreement within the experimental error.

The experimental values of the flux-averaged cross-sections $\langle{\sigma(E_{\rm{\gamma max}})}\rangle_{\rm{m}}$ for the $^{181}{\rm{Ta}}(\gamma,p)^{180\rm{m}}\rm{Hf}$ reaction were determined. The  $\langle{\sigma(E_{\rm{\gamma max}})}\rangle_{\rm{m}}$ values for the studied reaction at energy $E_{\rm{\gamma max}} >$ 55~MeV were obtained for the first time. 

The calculation of bremsstrahlung flux-averaged cross-section $\langle{\sigma(E_{\rm{\gamma max}})}\rangle_{{\rm th}}$ was carried out using the
cross-section values $\sigma(E)$ computed with the TALYS1.95 code for different level density models $LD$ 1–6. The theoretical estimates significantly underestimate both our experimental results and data from the literature \cite{27,28}. 
The obtained experimental $\langle{\sigma(E_{\rm{\gamma max}})}\rangle_{{\rm exp}}$ are closest to the theoretical calculation using the $LD$3 level density model -- the Generalized superfluid model.


A comparative analysis of the calculated total cross-sections for the reactions $^{181}{\rm{Ta}}(\gamma,p)^{180}\rm{Hf}$ and $^{181}{\rm{Ta}}(\gamma,n)^{180}\rm{Ta}$ was performed. The strength ratio $(\gamma,p)$ and $(\gamma,n)$ of the GDR photodisintegration components for the $^{181}\rm{Ta}$ nucleus was calculated using the isospin selection rules and equals to $9.81 \times 10^{-3}$. This value was compared with the ratio of weighted integrals, in which the total cross-sections for the $^{181}{\rm{Ta}}(\gamma,p)$ and $^{181}{\rm{Ta}}(\gamma,n)$ reactions obtained from the TALYS1.95 code with the $LD$3 model, were used. 

It was shown that the photoproton $(\gamma,p)$ to photoneutron $(\gamma,n)$ strength ratio of the GDR, found taking into account the experimental values for the reaction \\ $^{181}{\rm{Ta}}(\gamma,p)^{180\rm{m}}\rm{Hf}$, is consistent with the expected estimate according to the isospin selection rules, and the strength ratio obtained using the experimental cross-sections from the $(e,e'p)$ experiment. 

\section*{Acknowlegment}
The authors would like to thank the staff of the linear electron accelerator LUE-40 NSC KIPT, Kharkiv, Ukraine, for their cooperation in the realization of the experiment. 

This work was supported by the Slovak Research and Development Agency under No. APVV-20-0532, and the Slovak grant agency VEGA (Contract No. 2/0067/21). Funded by the EU NextGenerationEU through the Recovery and Resilience Plan for Slovakia under the project No. 09I03-03-V01-00069.

\section*{Declaration of competing interest}
The authors declare that they have no known competing financial interests or personal relationships that could have appeared to influence the work reported in this paper.

%
%

\end{document}